\documentstyle[12pt]{article}
\begin{document}
\title{Bounds for Turbulent Transport}
\author{Peter Constantin
\\Department  of Mathematics\\The University of Chicago }
\maketitle
\newtheorem{thm}{Theorem}
\newtheorem{prop}{Proposition}

We would like to pursue a mathematical approach to turbulence that
is able to predict bulk average quantities of experimental and 
practical significance. Among such bulk average quantities, the 
Nusselt number $N$, the enhanced bulk 
average heat transfer due to convection, is perhaps the simplest
physical objective of rigorous mathematical study of turbulence. 

The question we wish to address here is: what are
the optimal bounds on $N$ as a function of Rayleigh number $R$? What is the effect of the Prandtl number? 
Theoretical results (\cite{convection}, \cite{howard}, \cite{kr}) predict
a maximal ultimate behavior of $N\sim \sqrt{R}$.
The experimental (\cite{exp}) findings indicate however that
$$
N\sim R^q
$$
where the reported
values for $q$ belong approximately to the interval
$[\frac{2}{7}, \frac{1}{3}]$ for large $R$. 
The exponents $2/7$ and
$1/3$  have been discussed by several authors, 
(\cite{kad}, \cite{ss}). 
The recent results of (\cite{sreeni}) favor an exponent of  3/10 with 
logarithmic corrections.
We will describe below some very simple rigorous results.

The mathematical formulation of the problem is 
based on the three dimensional Boussinesq equations for
Rayleigh-B\'{e}nard convection (\cite{chandra}). These are  
a system of equations coupling the three dimensional Navier-Stokes equations 
\begin{equation}
\frac{\partial{\mathbf u}}{\partial t} +  {\mathbf u}\cdot\nabla {\mathbf u} + \nabla p = \sigma \Delta {\mathbf u} + \sigma RT{\mathbf e_z}, \quad\nabla\cdot {\mathbf u} = 0\label{ns}
\end{equation}
to a heat 
advection-diffusion equation
\begin{equation}
\frac {\partial T}{\partial t} + {\mathbf u}\cdot\nabla T
= \Delta T.\label{teq}
\end{equation}
The five unknowns, incompressible velocity, ${\mathbf u} = (u,v,w)$, pressure $p$, and temperature $T$ are functions of position
${\mathbf x} = (x,y,z)$ and time $t$. $R$ is the Rayleigh number and
$\sigma$ is the Prandtl number. ${\mathbf e_z}  = (0,0,1)$ is the unit vector in
the vertical direction.  For simplicity of exposition the equations have been non-dimensionalized: the vertical variable $z$ is scaled so that it belongs to the interval $[0,1]$, the horizontal independent variables $(x,y)$ belong to a square $Q\subset\mathbf{R}^2$ of side $L$.  The boundary conditions are as follows: all functions ($(u,v,w)$, $p$, $T$) are periodic in  $x$ and $y$ with period $L$; $u$, $v$, and $w$ vanish for $z =0,1$, and the temperature obeys 
$T=0$ at $z=1$, $T=1$ at $z=0$. We write
$$
\|f\|^2 = \frac{1}{L^2}\int_0^L \int_0^L \int_0^1|f(x,y,z)|^2\,dx\,dy\,dz
$$
for functions and vectors $f$. We use $<\cdots>$ for
long time average:
$$
\left<f\right> = \lim\sup_{t\to\infty}\frac{1}{t}\int_0^tf(s)ds.
$$
We write $\overline f$ the horizontal average
$$
\overline{f} = \frac{1}{L^{2}} \int_{0}^{L}\int_{0}^{L} f(x,y)\,dx\,dy.
$$
When a function depends on additional variables we write only 
the remaining variables after integration, so for instance
$$
\|\nabla {\mathbf u}(\cdot, t)\|^2 = \frac{1}{L^2}\int_0^L\int_0^L\int_0^1
|\nabla {\mathbf u}(x,y,z,t)|^2\,dx\,dy\,dz.
$$

The Nusselt number is given in terms of the long time average of the vertical heat flux:
\begin{equation}
N = 1 + \left\langle \int_{0}^{1} b(z) \, dz \right\rangle\label{nusselt}
\end{equation}
with
\begin{equation}
b(z,t) = \frac{1}{L^{2}} \int_{0}^{L}\int_{0}^{L}     w(x,y,z,t) T(x,y,z,t)\,dx\,dy = \overline{wT}(z,t).\label{bzt}
\end{equation}

A consequence of the equations of motion are two additional formulas
for the Nusselt number:
\begin{equation}
\left <\|\nabla T\|^2\right> = N \label{ngrt}
\end{equation}
and
\begin{equation}
\left <\|\nabla \mathbf{u}\|^2\right > = R(N-1)\label{ngru}.
\end{equation}

The classical result of Howard, conditioned on
on assumptions about statistical averages,
is that $N$ is bounded at very large Rayleigh numbers by
a multiple of $R^{\frac{1}{2}}$. The same kind of bound can be
derived without any conditions (\cite{dcon}). 
This bound is valid for all solutions, aspect ratios $L$ and
Prandtl numbers $\sigma$ and is also valid in a rotating frame, at arbitrary rotation speed.

The system is not isotropic: the direction of gravity is singled out.
Consider a function $\tau (z)$ that satisfies $\tau (0) = 1$, 
$\tau (1) = 0$, and express the temperature as 
\begin{equation}
T(x,y,z,t) = \tau (z) + \theta(x,y,z,t).
\end{equation}
The role of $\tau$ is that of a convenient background (\cite{let}, \cite{heat},
\cite{var} ) that carries the inhomogeneous boundary conditions; thus $\theta$ obeys the same homogeneous boundary conditions as the velocity. Note that, because $\tau $ does not depend on $x , y$
one has
\begin{equation}
T(x,y,z,t)-\overline{T}(z,t) =\theta(x,y,z,t) -\overline{\theta}(z,t).\label{tt}
\end{equation}

The equation obeyed by $\theta $ is 
\begin{equation}
\left (\partial _t + u\cdot\nabla -\Delta \right)\theta =
\tau^{\prime\prime} -w\tau^{\prime}\label{theteq}
\end{equation}
where we used $\tau^{\prime} =\frac{d\tau}{dz}$.
The horizontal average of the vertical velocity vanishes
identically because of incompressibility
$$
\overline {w}(z,t) = 0. 
$$
Therefore
the quantity $b(z,t)$ can be written as
\begin{equation}
b(z,t) = \frac{1}{L^{2}} \int_{0}^{L}\int_{0}^{L}     w(x,y,z,t) \left (\theta(x,y,z,t)- \overline{\theta}(z,t)\right )\,dx\,dy = \overline{w(T-\overline{T})}(z,t)\label{bz}
\end{equation}

Multiplying the equation (\ref{theteq}) by $\theta$ and integrating  
we obtain 
\begin{equation}
N + \left<\|\nabla \theta\|^2\right > = 2\left <-\int_0^1
  \tau^{\prime }(z)b(z)dz \right > +\int_0^1   \left (\tau^{\prime }(z)\right )^2dz.\label{theta}
\end{equation}
We will choose the background profile $\tau$ 
for simplicity to be 
a smooth profile concentrated in a boundary layer of width $\delta$,
$$
\tau (z) = P\left (z\over{\delta}\right )
$$
with $P(0) = 1$ and $P(s)= 0$ for $s\ge 1$.
Using only elementary facts (fundamental theorem of calculus, the boundary conditions  and the Schwartz
inequality) it is easy to see from (\ref{bz}) that
\begin{equation}
|b(z,t)| \le z\|\nabla {\mathbf {u}}(\cdot,t)\|\cdot\|\nabla (T-\overline{T})(\cdot,t)\|\label{una}
\end{equation}
holds for any $z$.
Let us  define
\begin{equation}
n = \left <\|\nabla (T-\overline{T})\|^2\right > \label{n}
\end{equation}
Note that, from the definition of $n$ and (\ref{ngrt}) it follows that
$$
n\le N.
$$
From (\ref{una}), (\ref{theta}) and (\ref{ngru}) we obtain
$$
N \le \frac{C}{\delta} + 2D\delta\sqrt{R(N-1)}\sqrt{n}
$$
with
\begin{equation}
C = \int_0^1\left (\frac{dP(s)}{ds}\right )^2ds\label{C}
\end{equation}
and 
\begin{equation}
D = \int_0^1s\left|\frac{dP(s)}{ds}\right |ds\label{D}.
\end{equation}
Optimizing in $\delta <1$ we get
$$
\delta^{-1} = \sqrt{{2D\over{C}}}\left\{R(N-1)n\right \}^{\frac{1}{4}} + 1
$$
and letting $P\to s$ we obtain
\begin{equation}
N\le 2(Rn)^{\frac{1}{4}}(N-1)^{\frac{1}{4}} + 1\label{interim}
\end{equation}
Thus we have proved 
\begin{thm}
Let
$$
n = \left <\|\nabla (T-\overline{T})\|^2\right >
$$
Then the Nusselt number (\ref{nusselt}, \ref{ngrt}, \ref{ngru})
for three dimensional Rayleigh-B\'{enard}
convection satisfies
$$
N \le 2^{\frac{4}{3}}\left (Rn\right )^{\frac{1}{3}} + 1.
$$
\end{thm}

If one has  no additional
information then, using $n\le N$ in the inequality above we obtain the
square-root bound
$$
N\le 4\sqrt{R} + 1
$$
(The prefactor is not optimal. The search for optimal prefactors is better
motivated for other systems, where the power law obtained rigorously coincides with the one observed in experiments. When that is the case then the
rigorous results can match experiments with remarkable accuracy (\cite{cd})). 
The exponent $1/3$ (or anything less than
$1/2$  for that matter) have not been proven rigorously for the
general system. The theorem above brings in the exponent 1/3 in the general
Boussinesq system conditionally, for slowly varying $n$. 
Another way by which the Nusselt number dependence on the Rayleigh number can be lower is if the Prandtl number is very high or in rotating (\cite{chp}) convection. If the Prandtl number is infinity then
the upper bound is closer to 1/3.
The equations of motion for infinite Prandtl number
Rayleigh-B\'{e}nard convection (\cite{chan}) are
\begin{equation}
- \Delta {\mathbf u} + \nabla p  = RT e_z,\quad \nabla\cdot{\mathbf u} = 0\label{nip}
\end{equation}
coupled with the advection-diffusion equation (\ref{teq}). Because (\ref{nip})
is time independent we say that $T$ obeys an active scalar equation.

An important observation, true even for the general
case (\ref{ns}) is that, in view of the boundary conditions
and incompressibility, not only the vertical component of velocity $w$ but also its normal derivative $\frac{\partial w}{\partial z}$ vanish at the vertical boundaries. 
Therefore
we can write
\begin{equation}
b(z,t) = \overline{\int_0^zdz_1\int_0^{z_1} 
\frac{\partial^2 w}{\partial z^2}(\cdot,s,t)ds\int_0^z
\left (\frac{\partial (T(\cdot, \sigma,t )- \overline{T}(\sigma,t))}{\partial z}\right ) d\sigma }. \label{basic}
\end{equation}
Consequently
\begin{equation}
|b(z,t)| \le z^{\frac{5}{2}}\left  \|\frac{\partial^2 w(\cdot, t)}{\partial z^2}\right \|_{L^{\infty}}\|\nabla (T-\overline{T})(\cdot, t)\|\label{bb}
\end{equation}
where $\|f\|_{L^{\infty}}$ is the sup-norm.
One can express 
$\frac{\partial^2w}{\partial z^2}$, in terms of $T-\overline{T}$. Indeed,
eliminating the pressure
from (\ref{nip}) one has
\begin{equation}
        \Delta^{2} w = 
        - R\Delta_{h} (T-\overline{T})     \label{bi}
\end{equation}
where $\Delta_h$ is  the Laplacian in the horizontal
directions $x$ and $y$. Using the boundary conditions we may write this as
\begin{equation}
        w = -R ( \Delta_{DN}^2 )^{-1} \Delta_{h} (T-\overline{T})  \label{eq:3}
\end{equation}
where $(\Delta_{DN}^2)^{-1}$ is the inverse bilaplacian with
homogeneous Dirichlet and Neumann boundary conditions.
Taking two $z$ derivatives then yields
\begin{equation}
   \frac{\partial^2w}{\partial z^2}     = -RB(T-\overline{T}) 
        \label{wzz}
\end{equation}
where the linear operator $B$ is given by
\begin{equation}
    B = \frac{\partial^2}{\partial z^2} ( \Delta_{DN}^{2} )^{-1} \Delta_{h}
    \label{B}
\end{equation}
The temperature equation obeys a maximum principle
so that
$$
0\le T\le 1
$$
holds pointwise in space and time. Therefore
$$
0\le |T-\overline{T}| \le 1
$$
holds. The operator $B$ is not  
bounded in $L^{\infty}$ but obeys a logarithmic extrapolation estimate.
This means that higher derivatives enter logarithmically in the bound; the estimate 
\begin{equation}
\left <\|B(T-\overline{T})\|_{L^{\infty}}^2\right >\le C_1^2
\left \{1+\log_+R\right\}^4
\label{est1}
\end{equation}
follows from the bounds in (\cite{nip}). The constant $C_1$ 
can be computed explicitly.
Therefore, using the same kind of background as above in (\ref{theta})
we obtain 
\begin{equation}
N\le \int_0^1\left (\tau^{\prime }(z)\right )^2dz + 2\int_0^1z^{\frac{5}{2}}
|\tau^{\prime}(z)|\left<\left\|\frac{\partial^2 w(\cdot,t)}{\partial z^2}\right \|_{L^{\infty}}\|\nabla(T-\overline{T})\|\right >dz
\label{maste}
\end{equation}
and consequently
\begin{equation}
N \le \frac{C}{\delta} + 2E\delta^{\frac{5}{2}}C_1R\left \{1+\log_+R\right\}^2
\sqrt{n}\label{interme}
\end{equation}
with $C$ defined in (\ref{C}), $C_1$ coming from (\ref{est1}) and
\begin{equation}
E = \int_0^1s^{\frac{5}{2}}P(s)ds.\label{E}
\end{equation}
optimizing in $\delta <1$ we find
$$
\delta^{-1} =\left [ \frac{5E}{C}C_1R\left \{1+\log_+R\right\}^2\sqrt{n}\right ]^{\frac{2}{7}} + 1
$$
and then letting $P\to s$ we deduce
\begin{thm}
    There exists a constant $C_2$ such that the Nusselt number for the 
    infinite Prandtl number equation is bounded by
$$
N \leq 1 + C_2R^{\frac{2}{7}}\left \{  1 + \log_{+}R\right 
    \}^{\frac{4}{7}}n^{\frac{1}{7}}. 
$$
\end{thm}

If no additional information is given then, using $n\le N$ in the
theorem above we recover the result
$$
N \leq 1 + C_2^{\frac{7}{6}}R^{\frac{1}{3}}\left \{  1 + \log_{+}R\right 
    \}^{\frac{2}{3}}
$$
of (\cite{nip}). The theorem brings in the exponent 2/7 as long as $n$ is not 
varying too much with $R$.

\begin{center}{\large \bf Discussion}\end{center}

\noindent There is no proof that the Nusselt number cannot ever scale
like $R^{1\over 2}$ for exceedingly large $R$. The experimental data 
do not seem however to encounter this behavior. Mathematically, the Nusselt
number represents the maximum (among all possible invariant measures) of the
expected value of the diameter of the global attractor in the energy 
dissipation norm. The functions on the attractor are not arbitrary, and 
may have certain properties that explain the experimentally observed 
bounds (\cite{heat}). In this paper we showed that if the ratio
$$
\frac{n}{N} = \frac{\left <\|\nabla (T -\overline{T})\|^2\right >}{\left <\|\nabla T\|^2\right>}
$$
is small then the Nusselt number dependence on Rayleigh is depleted.

\vspace{1cm}
\noindent{\bf Acknowledgment} This work was partially supported 
by NSF-DMS9802611 and by the ASCI Flash Center at the University of 
Chicago under DOE contract  B341495.

\end{document}